\documentstyle[12pt]{article} 
\font\tenrm=cmr10 

\newcommand{\bref}[1]{(\ref{#1})} 
\newcommand{\ct}[1]{\cite{#1}}

\newcommand{\be}{\begin{equation}} 
\newcommand{\ee}{\end{equation}} 

\begin{document}  
\titlepage  
  
\begin{flushright}  
{MPI-PhT/96-37\\}  
{CCNY-HEP-96/6\\}   
{CU-TP-801 \\ }
{hep-th/9709195\\}  
{December 1996\\}  
\end{flushright}  
\vglue 1cm

\begin{center}   
{ 
{\Large \bf An Approximate Large $N$ Method \\  
for Lattice Chiral Models    
\\}  
\vglue 1.0cm  
{Stuart Samuel\\}   
\bigskip  
{\it Max-Planck-Institut f\"ur Physik\\}
{\it Werner-Heisenberg-Institut\\} 
{\it F\"ohringer Ring 6\\} 
{\it 80805 Munich, Germany\\} 

\medskip 

{\it Department of Physics\\}
{\it Columbia University\\}
{\it New York, NY 10027, USA\\} 

\medskip 

{\it and \\} 

{\it Department of Physics$^{*}$\\}
{\it City College of New York\\}
{\it New York, NY 10031, USA\\} 
\vglue 0.8cm  
} 
 
\vglue 0.3cm  
  
{\bf Abstract} 
\end{center}  
{\rightskip=3pc\leftskip=3pc 
\quad An approximation is used  
that permits one to explicitly solve the 
two-point Schwinger-Dyson equations 
of the $U(N)$  lattice chiral models.  
The approximate solution correctly 
predicts a phase transition for dimensions $d$ 
greater than two.  
For $d \le 2 $, 
the system is in a single disordered phase 
with a mass gap.  
The method reproduces known $N=\infty$ results well 
for $d=1$. 
For $d=2$, there is a moderate difference with 
$N=\infty$ results 
only in the intermediate coupling constant region.  

}

\vfill
 
\textwidth 6.5truein
\hrule width 5.cm
\vskip 0.3truecm 
{\tenrm{
\noindent 
$^*$Permanent address.\\   
\hspace*{0.2cm}E-mail address: samuel@scisun.sci.ccny.cuny.edu\\}}
 
\newpage  
  
\baselineskip=20pt  
 
{\bf\large\noindent I.\ Introduction}  

The generation of a mass gap 
in two-dimensional spin systems 
is believed to be analogous to 
the generation of a non-zero string tension 
in four-dimensional non-abelian gauge theories 
\cite{migdal76,es79}.    
Of particular interest 
are the $d=2$ matrix chiral models.  
They are asymptotically free and have 
properties similar to $d=4$ gauge theories 
\cite{migdal76,es79,gs81}.  
A solution to the $N=\infty$ 
chiral models has not yet been found.  
In contrast, 
it is straightforward to solve 
the $O(N)$ vector spin models 
as $N \to \infty$.


The degrees of freedom of the $O(N)$ 
models are $N$-component vectors $v(x)$ assigned 
to the sites $x$ of a lattice
satisfying the constraint $v(x) \cdot v(x) =1$.  
If it were not for this constraint, 
the $O(N)$ vector model would be a theory 
of $N$ non-interacting particles.  
Thus the effect of the constraint is to introduce interactions. 
Due to these inter-particle forces, 
an interesting picture of mass generation arises 
\cite{gs81}.  
In terms of bare quanta, 
the interactions are repulsive 
and of order $1/N$.  
The quantum vacuum  
involves fluctuations 
in which any one of the $N$ quanta  
are created and destroyed.  
{}From a Euclidean viewpoint, 
such events correspond to particle loops, 
and so the vacuum is like a gas of closed loops.  
Such loops are small and dilute 
in strong couping 
since  
the mass of a particle is large.  
As one moves to weaker couplings, 
the mass decreases  
and the loops become larger and more plentiful.   
If interactions could be turned off, 
the mass would eventually vanish and a phase transition 
would occur.  
For dimensions $\le2$, 
two large closed loops generically intersect.  
Because there are repulsive forces at such intersection points, 
a particular loop feels a pressure 
from the gas of surrounding loops.  
For $d \le 2$ and $N \ge 3$, 
this pressure prevents any particular loop from becoming too large 
and generates an interaction-induced mass  
which never vanishes.  
By these means, 
the vector models avoid a phase transition when $d \le 2$.  
The models remain in the strong coupling phase 
in which the particles are massive and the system is disordered:   
Correlation functions fall off exponentially with distance.  
In the large $N$ limit, 
the vector model is exactly solvable:   
after interactions are incorporated into an effective coupling, 
particles are free.  
For $d \le 2$, 
the effective coupling never attains a value sufficient 
to produce a phase transition 
to a massless spontaneously broken phase.  

The picture for the matrix chiral models is 
qualitatively the same 
\cite{gs81}.  
Forces at the intersection of particle loops are repulsive, 
as can be seen from Figure 6 
of ref.\cite{gs81}.   
However, 
no effective free-theory formulation of the model 
has been derived, 
although   
an exact $S$-matrix has been proposed 
\cite{aas84,wiegmann84}.  
For $SU(N)$, 
the spectrum is 
$M_n = M sin \left( { n \pi/N} \right) / sin \left( { \pi/N} \right) $, 
for $n=1, 2, \dots , N-1$.  
Interestingly, 
as $N \to \infty$, 
the $S$-matrix becomes the identity matrix 
to leading order in a $1/N$ expansion.  
This suggests that perhaps the large $N$ limit 
of the $SU(N)$ chiral model is 
a free theory of particles.  

\medskip 
{\bf\large\noindent II.\ The Approximate Large $N$ Method}  

In this work, 
we obtain an approximate solution for the chiral models.  
Correlation functions and the mass gap are obtained.  
Because the method is 
similar to the one used in the vector models, 
it is useful to quickly review the vector models 
in the large $N$ limit.  
The action ${\cal A}$ 
for the $d$-dimensional $U (N)$ vector model is 
\be  
 {\cal A} = 
  \beta N\sum\limits_{x,\Delta } 
 {v^*\left( x \right)\cdot v\left( {x+\Delta } \right)} 
\quad , 
\label{eq1}  
\ee 
where $v(x)$ is an $N$-dimensional complex vector 
satisfying $v^*(x) \cdot v(x) = 1$, 
$\beta$ is proportional to the inverse coupling constant, 
and the sum over $\Delta$ is 
over the $2d$ nearest neighbor sites, 
i.e., 
$\Delta$ is $\pm e_1$, $\pm e_2$, $\dots$, $\pm e_d$, 
where $e_i$ is a unit vector in the $i$th direction.   
The $U(N)$ model is equivalent 
to the $O(2N)$ vector model.  

The lattice Schwinger-Dyson equation  
for the two-point function 
is  
$$ 
  \delta _{xy} = 
  \left\langle {v^*\left( x \right) \cdot 
   v\left( y \right)} \right\rangle - 
   \beta \sum\limits_\Delta  
   {\left\langle {v^*\left( x \right)\cdot 
   v\left( {y+\Delta } \right)} \right\rangle} \ + 
$$ 
\be
   \beta \sum\limits_\Delta  
   {\left\langle {v^*\left( x \right) \cdot 
    v\left( y \right)\,\,v^*\left( {y+\Delta } \right) \cdot 
   v\left( y \right)} \right\rangle} 
\quad .  
\label{eq2}  
\ee  
Since the four-point function enters in the last term, 
the vector model is not a free theory.  
However, 
large $N$ factorization allows the four-point function 
to be expressed 
in terms of two-point functions via 
$$ 
  \left\langle {v^*\left( x \right) 
   \cdot v\left( y \right)\,\,v^*\left( {y+\Delta } \right) 
   \cdot v\left( y \right)} \right\rangle =  
$$ 
\be
   \left\langle {v^*\left( x \right)\cdot 
   v\left( y \right)} \right\rangle 
  \left\langle 
   {v^*\left( {y+\Delta } \right)\cdot v\left( y \right)} 
   \right\rangle + O \left( 1 \over N \right)
\quad .  
\label{eq3}  
\ee 
The nearest neighbor expectation, 
$
 \left\langle 
   {v^*\left( {y+\Delta } \right)\cdot v\left( y \right)} 
   \right\rangle 
$,  
is independent of $y$.  
A linear equation for 
$
 \left\langle {v^*\left( x \right) \cdot 
   v\left( y \right)} \right\rangle 
$ 
is thus obtained 
when 
\bref{eq3} is substituted in 
\bref{eq2}.  
Consequently, 
the $N = \infty$ vector model is a free theory 
governed by an effective coupling 
\be
 \beta_{eff} = 
  { {\beta} \over {  
  1 + 2 d \beta 
  \left\langle {v^*\left( 0 \right) \cdot 
   v\left( \Delta \right)} \right\rangle 
               } }
\quad .  
\label{eq4}  
\ee  
For $d \le 2$, 
no phase transition is encountered as $\beta \to \infty$:   
The mass gap is a decreasing function of $\beta_{eff}$
but the effective coupling does not reach the critical value  
at which the mass gap vanishes.  
The theory remains in the strong-coupling disordered phase.  
For $d > 2$, 
a phase transition occurs to  
a weak-coupling spin-wave phase 
in which Goldstone bosons appear 
due to the breaking of the global $U(N)$ symmetry.  
This is in agreement with expectations 
\cite{kss83,gl84}. 

The action ${\cal A}$ of the matrix chiral model is 
\be 
 {\cal A} = 
  \beta N\sum\limits_{x,\Delta } 
   {Tr\left( {U^{\dag} \left( x \right) 
   U\left( {x+\Delta } \right) } \right)} 
\quad ,  
\label{eq5}  
\ee 
where $U\left( { x } \right)$ is a unitary matrix 
in a group $G$.  
The chiral model is the non-linear effective low-energy theory  
of the massless quark model involving $N$ flavors.  
For this reason when $d=4$, 
$G = SU(N)$ is often used with $N=2$ or $N=3$.  
The matrices $U(x)$ can be regarded as the Goldstone bosons 
of spontaneous 
broken chiral $SU_L (N) \times SU_R (N)$. 

In what follows, 
we treat the case $G=U(N)$.  
The Schwinger-Dyson equation 
for the two-point function is  
$$ 
  \delta _{xy} = 
  {1 \over N} \left\langle { 
    Tr\left( {U^{\dag} \left( x \right) 
  U\left( y \right)} \right) } \right\rangle - 
  \beta \sum\limits_\Delta  
   {1 \over N} \left\langle { 
   Tr\left( { U^{\dag} \left( x \right) 
  U\left( {y+\Delta } \right) } \right)  } \right\rangle + 
$$ 
\be
  \beta \sum\limits_\Delta  
  {1 \over N} \left\langle { 
   Tr\left( {U^{\dag} \left( x \right) 
  U\left( y \right)U^{\dag} \left( {y+\Delta } \right) 
   U\left( y \right)} \right)  } \right\rangle
\quad .  
\label{eq6}  
\ee 
Equation \bref{eq6} 
is not useful in weak coupling   
where $\beta$ becomes large  
because the last two terms  
individually become sizeable.  
Since the left-hand-side remains constant, 
a delicate cancellation between the two takes place.  

Equation \bref{eq6}  
is similar to 
eq.\bref{eq2}, 
except that matrix products arise 
instead of dot products.  
Consequently, 
the four-point function which enters in 
\bref{eq6} 
cannot be factorized in the large $N$ limit.  
However, 
one may consider the following vector-model-like replacement  
$$ 
  {1 \over N} \left\langle { 
    Tr\left( {U^{\dag} \left( x \right) 
   U\left( y \right)U^{\dag} \left( {y+\Delta } \right) 
   U\left( y \right)} \right) } \right\rangle 
     \to 
$$ 
\be 
  a{1 \over N} \left\langle {
   Tr\left( {U^{\dag} \left( x \right) 
   U\left( {y+\Delta } \right)} \right) } \right\rangle + 
   b{1 \over N} \left\langle { 
    Tr\left( {U^{\dag} \left( x \right) 
    U\left( y \right)} \right) } \right\rangle 
\quad ,  
\label{eq7}  
\ee 
where $a$ and $b$ are arbitrary functions of $\beta$, 
whose choices are at one's disposal.  

It is important to use values of $a$ and $b$ which 
reproduce the leading orders of perturbation theory, 
since one is interested in taking the continuum limit 
for which the coupling $g$ goes to zero.  
Writing 
$ 
  U\left( y \right) = 
    \exp \left[ {i\Phi \left( y \right)} \right] 
$  
and expanding about the identity matrix, 
one finds  
$
  U\left( y \right) U^{\dag} \left( {y+\Delta } \right) 
   U\left( y \right)=1+2i\Phi \left( y \right) - 
   i\Phi \left( {y+\Delta } \right) + \dots 
$. 
Hence, if 
$
  U\left( y \right) U^{\dag} \left( {y+\Delta } \right) 
   U\left( y \right)\to a U\left( {y+\Delta } \right) + 
   b U\left( y \right) 
$ 
then one needs, 
\be 
  a = -1 + O\left( {g^2} \right) 
\ , \quad \quad 
  b = 2 + O\left( {g^2} \right) 
\quad .  
\label{eq10}  
\ee 
Alternatively, 
one can perform an operator product expansion.  
Write 
$$ 
  U\left( y \right) U^{\dag} \left( {y+\Delta } \right) 
   U\left( y \right) = 
2U\left( y \right) - 
   U\left( {y+\Delta } \right) + 
$$ 
\be 
 \left( {U\left( {y+\Delta } \right) - 
   U\left( y \right)} \right)U^{\dag} \left( {y+\Delta } \right) 
    \left( {U\left( {y+\Delta } \right)-U\left( y \right)} \right) 
\quad .  
\label{eq11}  
\ee 
Note that the third term is a higher dimensional operator 
since it involves two derivatives.  
One again concludes that 
$a$ and $b$ have the perturbative expansions of 
eq.\bref{eq10}. 

After the substitution in 
eq.\bref{eq7} is made, 
one obtains 
\be 
{1 \over N} \left\langle { 
    Tr\left( {U^{\dag} \left( x \right) 
  U\left( y \right)} \right) } \right\rangle = 
  {{\delta _{xy}} \over {(1 + 2 d b \beta )}}  
   + 
  \beta_{eff} \sum\limits_\Delta  
   {1 \over N} \left\langle { 
   Tr\left( { U^{\dag} \left( x \right) 
  U\left( {y+\Delta } \right) } \right)  } \right\rangle 
\quad ,   
\label{eq12}  
\ee 
where 
\be 
  \beta_{eff} = { {\beta (1-a)} \over {1 + 2 d b \beta} }
\quad .  
\label{eq13}  
\ee 
One advantage of the substitution in
eq.\bref{eq7} 
is that 
the effective coupling $\beta_{eff}$ 
remains finite as $\beta \to \infty$ 
so that the corresponding equation 
for the two-point function does 
not involve the above-mentioned delicate 
cancellation between large terms.  

One possibility is to choose  
\be 
  a = -1 
\ , \quad \quad 
  b = 2 G_\Delta  
\quad ,   
\label{eq14}  
\ee 
where the average link $G_\Delta$ is 
\be 
  G_\Delta \equiv {1 \over N} 
   \left\langle { 
   Tr\left( {U^{\dag} \left( {y+\Delta } \right) 
   U\left( y \right)} \right) } \right\rangle 
\quad .  
\label{eq15}  
\ee 
The choice 
in eq.\bref{eq14}  
produces the correct result 
in eq.\bref{eq7} 
when $y=x$.  
As a side remark, 
selecting $a=0$ and $b = G_\Delta$ 
yields the two-point function of the vector model 
in the large $N$ limit.  

The the equation for two-point function 
$ 
  G\left( {x,y} \right) \equiv 
  \left\langle {Tr\left( {U^{\dag} \left( x \right) 
    U\left( y \right)} \right)} \right\rangle / N 
$ 
in eq.\bref{eq12}  
is linear.   
The solution is 
\be 
  G\left( {x,y} \right) = 
  {1 \over {1+2d\beta b}}\int\limits_{-\pi }^\pi  
   {{{d^dp} \over {\left( {2\pi } \right)^d}}} 
   {{\exp \left[ {ip\cdot \left( {y-x} \right)} \right]} 
     \over {1-2\beta_{eff} 
     \sum\limits_{i=1}^d {\cos \left( {p_i} \right)}}} 
\quad .  
\label{eq16}  
\ee 
If one requires that $G\left( {x,y} \right) =  1$ for $y=x$,  
then 
the self-consistency condition 
\be 
  1 = {1 \over {1+2d\beta b}} 
   \int\limits_{-\pi }^\pi  
    {{{d^dp} \over {\left( {2\pi } \right)^d}}} 
   {1 \over {1-2\beta _{eff}\sum\limits_{i=1}^d 
     {\cos \left( {p_i} \right)}}} 
\quad   
\label{eq17}  
\ee 
must be satisfied.  

A more detailed analysis reveals that 
the substitution in
eq.\bref{eq7} 
with $a$ and $b$ given in 
eq.\bref{eq14} 
is guaranteed to reproduce the $O (1)$ and 
$O \left( {g^2} \right)$ terms of correlation functions 
when a perturbative expansion of the theory is performed.  
Furthermore, 
eqs.\bref{eq14} and \bref{eq16} 
lead to qualitatively 
the same results as in the vector model case: 
For $d \le 2$, 
the chiral model is in the disordered phase and 
there is a mass gap for all values of $\beta$. 
For $d > 2$, 
a transition occurs at a finite value of $\beta$
to a phase in which global symmetries 
are spontaneous broken.  
This is in agreement with expectations 
\cite{migdal76,es79,gs81}. 

\medskip 
{\bf\large\noindent III.\ The One-Dimensional Case}  

When $d=1$, 
the chiral model can be solved exactly. 
Solutions are known for the continuum and lattice cases at finite $N$ 
and for $N=\infty$ 
\cite{gw80,wadia79}.  
This case allows one to test the approximate large $N$ method.  
When  $d=1$, 
solving eqs.\bref{eq7} and \bref{eq16} 
gives  
\be 
  G\left( {x,y} \right) = 
  \left( {{{1-\sqrt {1-4\lambda^2}} \over {2\lambda }} 
   } \right)^{\left| y - x \right|} 
\quad ,   
\label{eq18}  
\ee 
where 
\be 
  \lambda = 
  {{\left( {1-a} \right)\beta } \over {1+2\beta b}} 
   = \beta_{eff} 
\quad .  
\label{eq19}  
\ee 
As a consequence 
of the self-consistency condition 
in eq.\bref{eq17}, 
one also has 
\be 
  b = {{-1+\sqrt {1+4\beta ^2 
   \left( {1-a} \right)^2}} \over {2\beta }} 
\quad .  
\label{eq20}  
\ee 
For large $\beta$, 
the mass gap $m$ is 
\be 
  m \Lambda^{-1} = 
  {1 \over {2\left( {1-a} \right)\beta }} - 
  {1 \over {48\beta ^2\left( {1-a} \right)^2}} + \dots  
\quad ,  
\label{eq21}  
\ee 
where $\Lambda^{-1}$ is the lattice spacing, 
which is often denoted by $a$.  

The continuum limit is obtained by taking 
$ 
  \beta \to \infty \ 
$ 
and 
$ 
\Lambda^{-1} \to 0  
$ 
with 
$ 
  \beta \Lambda = {1 / \left( {g_c^2 N} \right) } 
$ 
fixed.  
Here, $g_c$ is the continuum coupling constant. 
Hence, 
\be 
  m = {{g_c^2 N} \over 4} 
\quad .  
\label{eq22}  
\ee 
This is the exact $d=1$ result for the $U(N)$ gauge theory.  

\medskip 
{\bf\large\noindent IV.\ The Two-Dimensional Case}  

When $d=2$, 
the approximate large $N$ method leads to 
a mass gap which is exponentially small as $\beta \to \infty$:  
\be 
  m\Lambda ^{-1}\approx  
  \exp \left[ {-2\pi \beta \left( {1-a} \right)} \right] 
\quad .  
\label{eq23}  
\ee 
For $a=-1$, 
$
  m \approx 
  \Lambda \exp \left[ {-4\pi \beta } \right] 
$.  
It is remarkable that an exponentially suppressed mass gap 
is obtained, 
given that the substitution in 
eq.\bref{eq7} 
is guaranteed only to reproduce perturbative results 
to order $g^2$.  
However, the mass gap 
in eq.\bref{eq23} 
for $a \to -1$ 
is not as suppressed as the result 
obtained from  perturbative renormalization group analysis, 
which gives 
$ 
  m \approx 
  \Lambda \exp \left[ {-8\pi \beta } \right] 
$ 
\cite{ms80}.   

One possibility to overcome this discrepancy is to permit 
the coefficients $a$ and $b$ in 
eq.\bref{eq7} 
to depend on the distance between $x$ and $y$. 
For weak coupling, 
$a \sim -1$ at short distances
but at large distances 
$a \sim -3$.  
However, 
one would have to find some theoretical justification 
for this dependence of $a$ on $\left| y - x \right|$. 

\medskip 
{\bf\large\noindent 
V.\ The Matching of Both Strong and Weak Coupling}  

Although the values of $a$ and $b$ 
in eq.\bref{eq14} 
produce good results 
for weak couplings, 
they fail to do so in the strong coupling region:   
For example, 
the mass gap behaves as $-ln(2\beta) + O( \beta)$ 
instead of $-ln(\beta) + O( \beta)$.  
To obtain results which are accurate in both strong 
and weak coupling, 
we have found that 
\be 
  a = -\left({  G_\Delta }\right)^2
\ , \quad \quad 
  b =  G_\Delta + \left({  G_\Delta }\right)^3 
\quad   
\label{eq25}  
\ee 
works reasonably well. 
These values of $a$ and $b$ satisfy 
the weak coupling limits in 
eq.\bref{eq10} 
since 
$G_\Delta = 1 - g^2 N/(4d) + O (g^4)$.   
In addition, 
one can show that they reproduce 
correlation functions correctly to $O \left( {g^2} \right)$ 
in a perturbative expansion. 
It turns out that 
eq.\bref{eq25} 
produces the correct $N=\infty$ strong coupling expansions  
for the mass gap and $G_\Delta$ to order $\beta^3$.  
Since 
eqs.\bref{eq16} and \bref{eq25} 
give good results in both the strong and weak coupling limits, 
there is a reasonable chance that results are good 
throughout the entire coupling constant region.  
The idea of weak--strong coupling interpolation  
recently was used successfully 
in an approximate analytic computation 
of the $0^{++}$ glueball mass 
in the $SU(3)$ gauge theory 
\cite{samuel96}. 

To examine how well 
eq.\bref{eq25} 
works for intermediate couplings, 
one can used the solvable $d=1$ case 
as a guide.  
When $N=\infty$, 
$G_\Delta = \beta$ for $\beta \le 0.5$ 
and $G_\Delta = 1- 1/(4\beta)$ for $\beta \ge 0.5$ 
\cite{gw80,wadia79}, 
and the two-point function is given by 
\be 
    G\left( {x,y} \right) = 
  \left( { G_\Delta } \right)^{\left| y - x \right|} 
\quad .  
\label{eq26}  
\ee   

Using the approximate large $N$ method 
with 
eqs.\bref{eq20} and \bref{eq25}, 
one obtains the following equation for $G_\Delta$: 
\be 
  \left( { G_\Delta } \right)^4 + 
   { {G_\Delta} \over {\beta} } = 1
\quad .  
\label{eq27}  
\ee 
For the approximate large $N$ method, 
the two-point correlation function is again 
given by 
eq.\bref{eq26} 
but with $G_\Delta$ determined from 
eq.\bref{eq27}.  
Hence the difference between the exact and approximate results 
is governed by $G_\Delta$.  
Figure 1 displays $G_\Delta$.  
The approximate large $N$ method yields values of $G_\Delta$ 
which are slightly less than the exact $N=\infty$ value. 
The difference reaches a maximum of $ 6.7 $\% near $\beta = 0.6$.  
Very good agreement is, of course, obtained for $\beta$ small 
or large.  

The lattice $N=\infty$ $d=2$ chiral model 
has not been exactly solved, 
but Monte Carlo data is available for finite values of $N$ 
and 
there exist analytic weak and strong coupling results 
\cite{rv94a,rv94b,crv95}.   
In Figure 2, 
we compare the approximate large $N$ method 
for the values of $a$ and $b$   
in eq.\bref{eq25} 
versus weak and strong coupling results. 
For the average link $G_\Delta$, 
displayed in Figure 2a, 
the approximate method agrees well 
with the $11$th order $N=\infty$ strong coupling series of 
ref.\cite{gs81} 
for $\beta < 0.25$.  
It also agrees the weak coupling $N=\infty$ series, 
$1 - 1/(8\beta) - 1/(256 \beta^2) + \dots$ for large $\beta$, 
but in the intermediate coupling region it is below 
both curves and below the $U(2)$ Monte Carlo data 
\cite{samuel81}. 
Since the data for $U(2)$ does not differ greatly from 
the data for $SU(10)$ of 
\ct{crv95}, 
the dots in Figure 2a are probably close to the $N=\infty$ results. 
By allowing $a$ and $b$ to be a polynomial in $G_\Delta$ 
of sufficient high order, 
the approximate large N method can be adjusted to 
produce more terms in the strong coupling series.  
When this is done, 
agreement for the average link 
is obtained to less than $5$\% 
throughout the entire coupling constant region.   

{}From the mass gap, 
$ m \Lambda^{-1} \equiv f (g)$,  
one can obtain a lattice beta function $\beta (g)$ 
(not to be confused with the inverse coupling) via 
$ 
  -{ {\beta (g) }/ {g} } = 
{ {f} / 
     {   \left( g { {\partial f} \over  {\partial g}  } \right) } }
$.  
In Figure 2a, 
the approximate method is compared 
to the strong coupling result 
\cite{gs81}  
and to perturbation theory for which 
$- \beta (g)/g = g^2 N / (16 \pi) + \dots$.  
Excellent agreement is obtained for $\beta < 0.25$.  
In weak coupling, 
the approximate large N method gives results for 
$- \beta (g)/g$ which are twice as large because 
the mass gap is exponentially suppressed only 
by half as much as the renormalization group result.  

\medskip 
{\bf\large\noindent VI.\ Conclusion}  

In this work, 
we have used a vector-model-like method 
to linearized the lattice Schwinger-Dyson two-point function 
in the $U(N)$ chiral model.  
For $d \le 2$, 
the approximate method 
produces no phase transition.  
This is in agreement with expectations 
for the finite $N$ case: 
The system is believed to be in the disordered phase 
and to have a mass gap for all couplings.  
For $N = \infty$, 
a phase transition occurs 
\ct{gw80,wadia79,gs80},  
even though the system is disordered 
on both sides of the transition.  
For $d \le 2$, 
our approximate method  
does not see this large $N$ transition,  
probably because 
the transition is rather mild and because it occurs 
in the intermediate coupling constant region, 
where the method is least accurate.   
For $d > 2$, 
a phase transition occurs for all $N$ 
and is expected to be first order 
\ct{gl84,pw84}.  
The approximate method correctly sees the transition
and correctly predicts the nature of the phases: 
symmetry breaking with massless 
Goldstone bosons at weaking coupling, 
while a disordered phase with a mass gap in strong coupling. 
Thus the method gives 
what-is-believed-to-be the correct qualitative phase diagram
in all dimensions 
for the finite $N$ case.  
By adjusting the method to agree with strong coupling expansions, 
the approximate method gives reasonably good results 
throughout the entire coupling constant regime 
for average link and mass gap. 

It is of interest to adapt the method to lattice gauge theories.  
The Schwinger-Dyson equations 
lead to relations among Wilson loops 
\cite{foerster79,eguchi79,mm79}.  
A non-self-intersecting loop is related to loops modified 
at a link 
by the addition of plaquettes in the $2d$ directions 
minus loops modified by 
the addition of $2d$ ``twisted plaquettes.''  
The structure of the equation is similar to 
eq.\bref{eq6} 
in the sense 
that the inverse coupling $\beta$ multiplies the two terms 
and there is a sizeable cancellation between the two terms 
in weak coupling.  
The analog approximate large $N$ method involves replacing 
each twisted-plaquette term by an untwisted loop 
and an original loop multiplied respectively 
by coefficients $a$ and $b$.  
We have three comments on this approximate approach 
to large $N$  
for lattice gauge theories.  
(i) Although Wilson loops are related to Wilson loops 
after the substitution 
of the twisted-plaquette terms is made,
there appears to be no way to find an analytic expression 
for Wilson loops when $d \ne 2$.  
This problem is probably related to the lack of an analytic description 
of free lattice string theory.  
(ii) When $d>2$, 
the method appears to depend on the link on which 
the Schwinger-Dyson equations are derived.  
This implies that certain consistency issues must be addressed.  
(iii) When $d=2$, 
the approximation can be consistently applied 
to the lattice gauge theory.  
For a $U(N)$ gauge group,   
the results are identical to those of the $U(N)$ chiral model.  
The value of a non-self-intersecting Wilson loop $WL$ 
of area $A$ is 
\be 
 WL = 
  \left( {{{1-\sqrt {1-4\lambda^2}} \over {2\lambda }} 
   } \right)^{A} 
\quad ,  
\label{eq29}  
\ee 
where $\lambda$ is given in 
eq.\bref{eq19}. 
The choice $a = - \left( { \langle Tr U_P \rangle /N } \right)^2$ 
and 
$ 
 b = \left( { \langle Tr U_P \rangle /N } \right) + 
  \left( { \langle Tr U_P \rangle /N } \right)^3 
$, 
where $\langle Tr U_P \rangle /N$ is the average plaquette 
(these values of $a$ and $b$ are the gauge-theory analog of 
eq.\bref{eq25}),  
leads to good results. 
The self-consistency condition becomes 
$ 
   \left( {  \langle Tr U_P \rangle /N } \right)^4 + 
   { { \langle Tr U_P \rangle / (N\beta) }  } = 1
$.  
The value of the Wilson loop then becomes 
\be 
  \left( {  \langle Tr U_P \rangle /N } \right)^{A} 
\quad .  
\label{eq30}  
\ee 
In the exact large $N$ limit, 
the Wilson loop value is given by 
eq.\bref{eq30} 
with $ \langle Tr U_P \rangle /N = \beta $ 
for $\beta \le 0.5$ 
and $ \langle Tr U_P \rangle /N = 1-1/(4\beta) $ 
for $\beta \ge 0.5$.  
Thus the discrepancy between the approximate and exact 
large $N$ results is the same as in the $d=1$ chiral model: 
The error is at most $6.7$\% near $\beta = 0.6$ 
and there is good agreement in the strong and weak coupling regimes.  
When a continuum limit is taken, 
the exact continuum solution is obtained.


The picture of mass-gap generation for spin models in $d \le 2$ 
has an analogy for gauge models in $d \le 4$.  
In the lattice gauge theory, 
the strong coupling expansion 
involves sums over closed surfaces.  
{}From a Euclidean viewpoint, 
such surfaces can be thought of as 
the propagation of strings.  
When two surfaces overlap, 
there are repulsive forces.  
In fact, 
the surface-surface interactions involve potentials 
which are exactly the same as in the chiral-model case 
\cite{gs81}.  
The quantum vacuum  
involves fluctuations 
in which strings appear, expand, contract 
and self-annihilate.    
This creates a medium 
in which any individual string must propagate.  
When $d \le 4$, 
two large surfaces generically intersect, 
and since forces are repulsive, 
the medium creates a pressure 
which inhibits strings from becoming large. 
If this mechanism is sufficient robust 
as $\beta$ increases and $g^2$ becomes smaller, 
then the string tension will remain non-zero 
in the continuum limit.  
The same mechanism can inhibit 
the spanning surface of a Wilson-loop computation 
from becoming large.  
Hence one arrives at a possible physical picture 
for confinement and mass-gap generation 
in $d \le 4 $ gauge theories.  
The idea fails for $d > 4$ 
because two two-dimensional surfaces 
do not generally intersect.  

\medskip 
{\bf\large\noindent Acknowledgments}  

I thank Jan Plefka for discussions.  
This work was supported in part 
by the Humboldt Foundation and 
by the National Science Foundation under the grant  
(PHY-9420615).  

\medskip 


\medskip   
{\bf\large\noindent Figure Captions}  
\medskip 

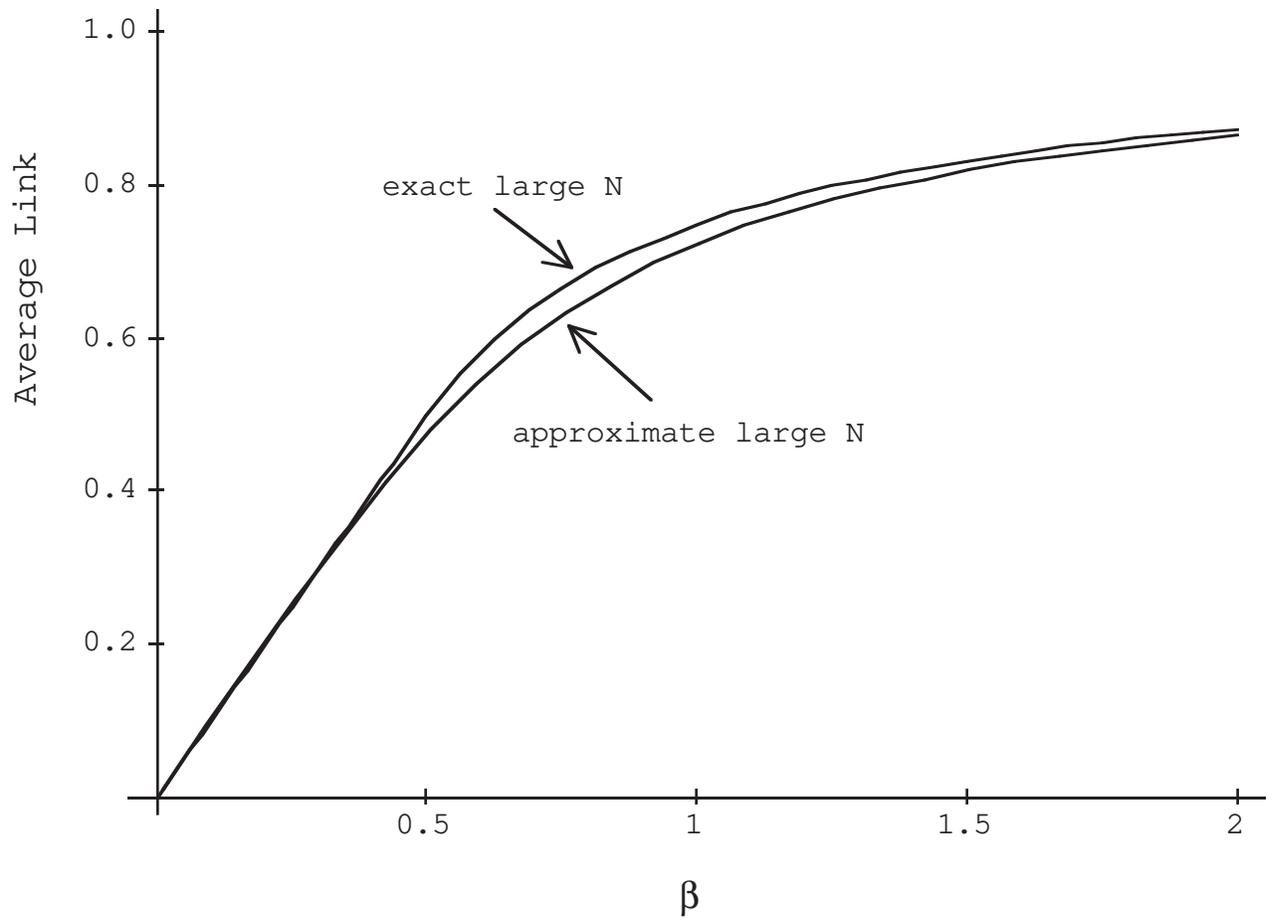
Figure 1. The Average Link $G_\Delta$ Versus $\beta$ 
for the $N=\infty$ $d=1$ Case. 

\medskip 

Figure 2a. The Average Link $G_\Delta$ Versus $\beta$ 
for the $N=\infty$ $d=2$ Case.  
The dots are $N=2$ Monte Carlo data.  

\medskip 

Figure 2b. The Beta Function 
$ -{ {\beta (g) } / {g} }$ 
Versus the Inverse Coupling $\beta$.  
The $N=\infty$ perturbative term comes
from renormalization group analysis.  

\vfil 

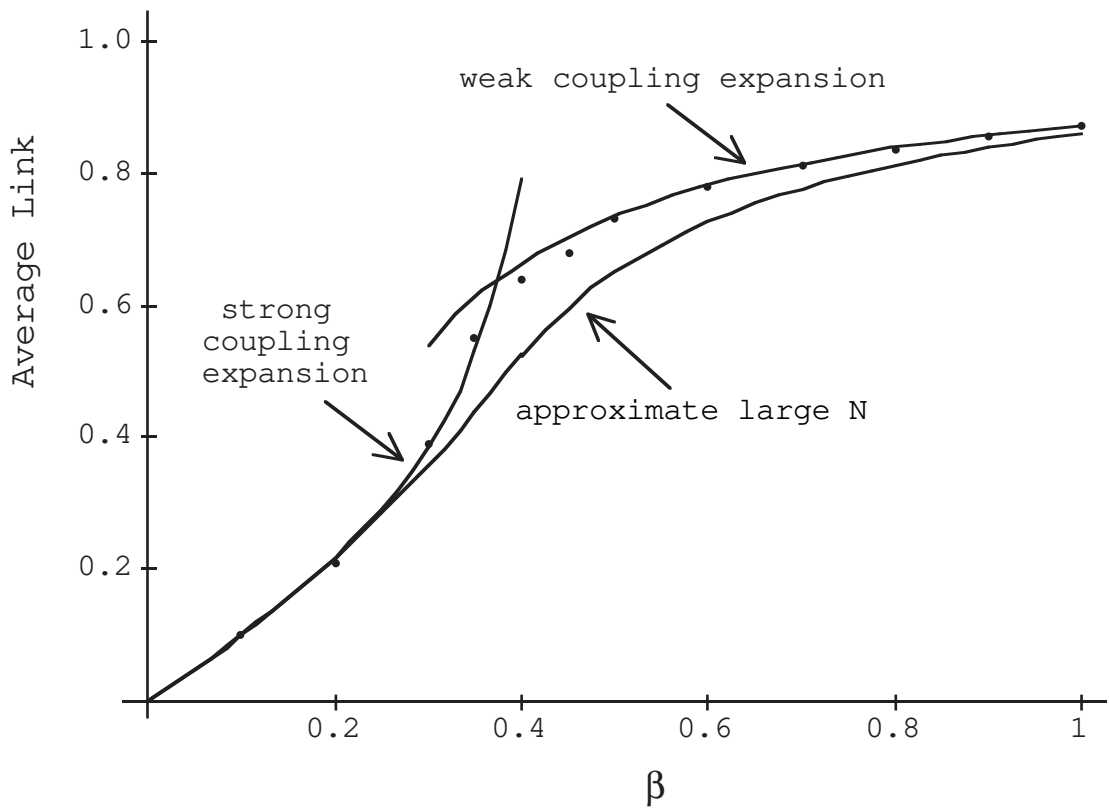
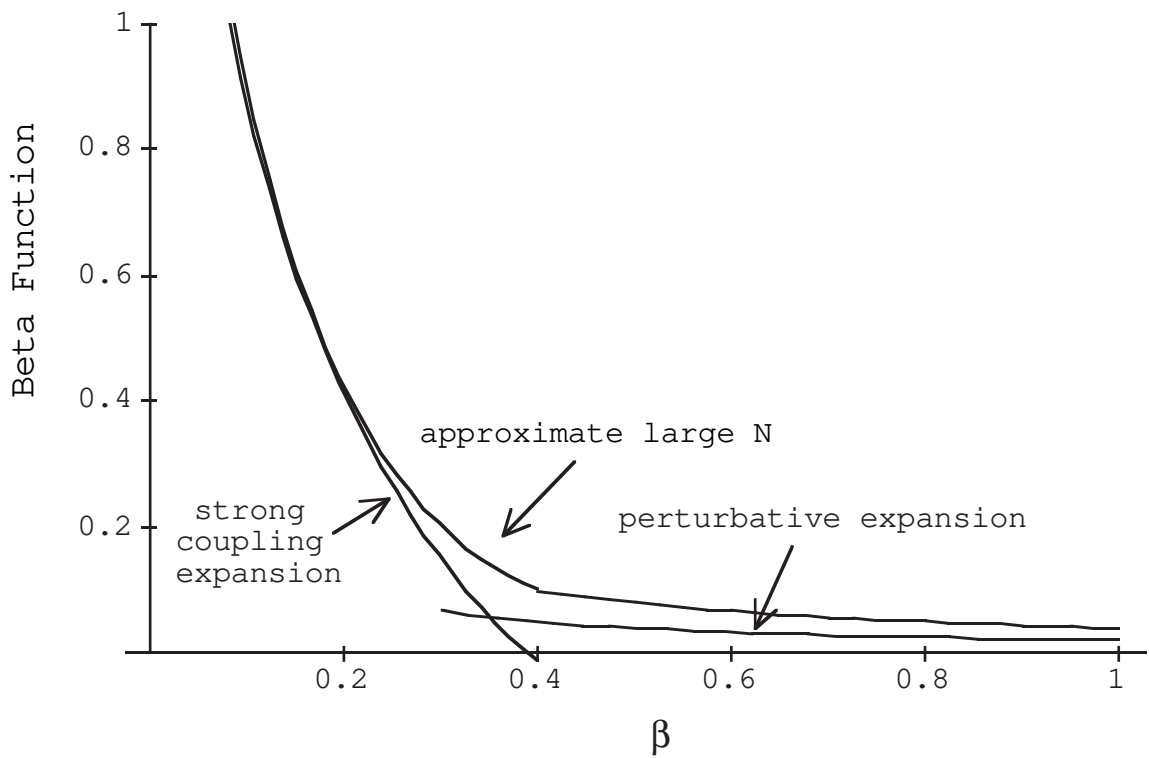
\end{document}